\documentclass[11pt]{article}
\usepackage[a4paper,margin=1in]{geometry}
\usepackage{microtype}
\usepackage{booktabs}
\usepackage{longtable}
\usepackage{array}
\usepackage{xurl}
\usepackage[colorlinks=true,linkcolor=blue,citecolor=blue,urlcolor=blue]{hyperref}

\newcolumntype{L}[1]{>{\raggedright\arraybackslash}p{#1}}
\title{A Practice Auditing Framework for Large Language Model Use: Collective Empiricism, Pseudo-Rational Cognition, and Governance of AI-Generated Content}
\author{Yang Zhao\\\texttt{zyxiaoqi7@gmail.com}\and Yingshuo Li\\\texttt{lys20020624@gmail.com}\and Zeyu Zhang\thanks{Corresponding author}\\School of Information and Electronics, Beijing Institute of Technology\\\texttt{zhangzeyu@bit.edu.cn}}
\date{}

\begin{document}
\maketitle

\begin{abstract}
Large language models are increasingly used for knowledge acquisition, code generation, image creation, academic writing, and agent-based systems. In these settings, users can obtain highly structured answers, plans, and judgments without sufficient domain practice. However, structured output does not mean that the user has completed a rational understanding of the concrete problem. This paper addresses LLM interaction and governance of AI-generated content by proposing a practice auditing framework. The framework explains how AI outputs may be transformed from compressed collective experience into user-level pseudo-rational cognition, and how unaudited AI-generated content may create cognitive loops and governance risks when it enters future contexts, memory systems, retrieval spaces, or detection mechanisms.

The paper first introduces the concept of collective empiricism to describe the mechanism by which large language models, without subject-centered practice, statistically learn, semantically reorganize, and contextually adapt large-scale human experience materials into outputs that appear empirical and rational. It then defines pseudo-rational cognition as a condition in which users mistake AI-generated structured expression for their own rational understanding despite lacking sufficient sensory materials, practice feedback, and conditional analysis. On this basis, the paper analyzes the illusion of AI subjectivity, subjectivity structures embedded in input materials, template loops in AI-AI conversations, statistical misjudgment in AIGC detection, and memory pollution or self-reinforcement caused by AI-generated content entering future retrieval spaces.

To mitigate these risks, the paper proposes a practice auditing process for AI use: requirement definition, problem-boundary identification, evidence-source auditing, practical validation, reverse questioning, logging, version management, rollback mechanisms, and renewed cognition. The framework does not deny the productivity value of AI. Instead, it argues that LLM outputs should be returned to verifiable, reproducible, and intervenable processes of practice. The contribution of this paper is to provide an analyzable and auditable theoretical framework for cognitive risks in LLM use, AI-generated content governance, long-term memory systems, and human-AI interaction.
\end{abstract}


\section{Introduction}

Large language models have increasingly become general-purpose interfaces for knowledge acquisition, code generation, image creation, academic writing, and agent-based automation. In these settings, users no longer interact with AI systems merely as search tools, but as systems that generate structured plans, explanations, judgments, code, summaries, and reusable intermediate artifacts. These outputs may further enter future prompts, memory modules, retrieval-augmented generation pipelines, AI-agent skill libraries, or AIGC detection workflows. Therefore, the key problem is no longer only whether an AI output is factually correct in a single turn, but how AI-generated structured content participates in the user's longer process of understanding, decision-making, validation, and revision.

This shift creates a governance problem. A user who does not understand the technical structure behind an artifact may still generate code, images, reports, or system designs that appear complete. A running program does not by itself show that the user understands dependency management, exception handling, maintenance, deployment risk, or long-term feedback. A visually appealing generated image does not show that the user understands composition, style transfer, denoising, model bias, or aesthetic judgment. In such cases, the model can compress collective experience into usable outputs, but the user may not have passed through the corresponding process of sensory accumulation, practical validation, and rational abstraction.

The purpose of this paper is not to reject AI use, nor to morally criticize users. Instead, it develops a practice auditing framework for LLM use and AI-generated content governance. The framework explains the relation among model outputs, user cognition, retrieval-space pollution, long-term memory, and practical validation, while retaining materialist epistemology as the philosophical source of its audit logic.

\subsection{Contributions}

The main contributions of this paper are as follows:

\begin{enumerate}
\item It introduces collective empiricism as a concept for describing how large language models compress, reorganize, and contextually adapt large-scale human experience materials without subject-centered practice.
\item It introduces pseudo-rational cognition as a concept for analyzing how users may mistake AI-generated structured expression for their own rational understanding when practice validation and conditional analysis are absent.
\item It analyzes loop risks that may emerge when AI-generated content enters future contexts, long-term memory, retrieval spaces, AI-AI conversations, and AIGC detection mechanisms, including template loops, memory pollution, skill debt, and statistical misjudgment.
\item It proposes a practice auditing framework for LLM use, emphasizing requirement definition, evidence auditing, practical validation, reverse questioning, logging, version management, and rollback mechanisms to support verifiable, reproducible, and intervenable AI use.
\end{enumerate}

\section{Theoretical Background: Materialist Epistemology and LLM Use}

The problem of AI use appears to be a technical problem, but it is also an epistemological problem. When users ask an AI system a question, receive a structured answer, and use that answer for creation, engineering, judgment, decision-making, or expression, they have already entered a new kind of cognitive activity. The difficulty is that this activity does not automatically mean that the user has completed a rational understanding of the object at hand.

In a traditional process of cognition, a person usually contacts an object, gathers sensory materials, discovers contradictions, performs abstraction, forms judgments, returns to practice for verification, and continuously revises their understanding. LLM use compresses this process. Users may obtain highly structured expressions without sufficient sensory accumulation. These expressions often look like rational cognition: they contain concepts, classifications, reasoning, steps, and conclusions. Yet they may not have passed through the user's own practice.

Thus, the central issue in the AI era is not simply whether an AI system can produce an answer. It is how users understand the place of that answer in their own cognitive process. If users treat AI-generated structured expression as though it were their own completed understanding, a displacement of cognition occurs. AI output may serve as material, tool, problem map, or preliminary plan, but it cannot directly replace the user's understanding of concrete objects, concrete contradictions, and concrete conditions.

From materialist epistemology, this paper argues that AI use should be returned to the process of practice, cognition, renewed practice, and renewed cognition \cite{mao1937practice}. AI can increase the efficiency of cognition, but it cannot cancel the moment of practice. It can expand the sources of experience, but it cannot automatically digest that experience. It can generate rational expression, but it cannot guarantee that such expression has become the user's own rational cognition.

\subsection{From Practice to Cognition}

Materialist epistemology holds that human cognition originates from practice and develops through practice \cite{mao1937practice}. Cognition does not arise from nothing, nor does it naturally grow out of linguistic structure alone. It is gradually formed in real activity, through contacting objects, handling problems, encountering limits, forming judgments, receiving feedback, and correcting errors.

The sequence of practice, cognition, renewed practice, and renewed cognition is not an abstract slogan. It is a basic path by which people obtain reliable understanding in reality. Initial practice provides sensory materials, through which people discover phenomena, differences, resistances, and contradictions. They then analyze, compare, generalize, and abstract these materials into preliminary rational cognition. But this cognition is not automatically correct. It must return to practice for verification. Only after practical testing, failure correction, and conditional analysis can cognition gradually approach truth under concrete conditions.

AI changes the speed of this process, but it does not change its basic law. A user can obtain a complete plan from a single prompt, or acquire many terms, frameworks, and judgments through repeated prompting. But these contents do not become the user's own cognition simply because they are complete in expression. What the user receives is an external expression compressed, reorganized, and generated by AI, not a practice process completed by the user.

For example, a user can ask AI to generate a small program. The program may run, the interface may be interactive, and its functions may look similar to those produced by an experienced developer. But this does not mean that the user understands code structure, exception handling, dependency relations, version management, security risks, maintenance, or user feedback. Running code is only an initial phenomenon. Whether the project can be maintained, extended, deployed, and debugged over time is the real question of practical cognition.

Similarly, a user can ask a multimodal model to generate a stylized image. A good-looking image does not mean that the user understands composition, style transfer, noise generation, denoising, model bias, or aesthetic judgment. The user has acquired descriptive ability, or prompt-expression ability, not a complete capacity for artistic practice.

This does not deny the value of AI-assisted creation. On the contrary, AI matters precisely because it lowers the threshold for participation in creation, engineering, and expression. But lowering the threshold does not erase differences in capability. AI can help users cross the starting line, but it cannot automatically supply the problem judgment, maintenance ability, and responsibility that are formed through long-term practice.

Therefore, the epistemological question in AI use is not whether users use AI, but whether they return AI output to practice for verification. Real AI-use capability is not merely receiving an answer after asking a question. It is the ability to judge under what conditions the answer holds, under what conditions it fails, and how it should be revised through practical feedback.

\subsection{Sensory Cognition and Rational Cognition}

Sensory cognition consists of materials obtained through direct contact with objects in practice, including phenomena, experience, failure, feedback, limits, and concrete conditions. Rational cognition consists of concepts, logic, theories, and methods formed through abstraction, synthesis, and judgment based on sensory materials. These two forms are not separate; they form a continuous relation of development.

Without sensory cognition, rational cognition easily becomes an empty stack of concepts. Without rational cognition, sensory cognition easily remains scattered experience and intuitive judgment. Reliable cognition must move back and forth between sensory material and rational abstraction.

The problem in the AI era is that users can bypass large amounts of sensory accumulation and directly obtain the form of rational expression. A person without long-term coding experience can obtain architecture diagrams, module breakdowns, interface designs, and deployment suggestions. A person without long-term research in a domain can obtain terminology, paper lineages, method classifications, and critical frameworks. A person without experience maintaining real products can ask AI to generate project descriptions, roadmaps, and business analyses.

These expressions are not necessarily wrong; often they are useful. The problem is that users may mistake the acquisition of rational expression for the completion of rational cognition. Real rational cognition is not the ability to recite a set of terms. It is the ability to identify principal and secondary contradictions in concrete problems, distinguish common conditions from special conditions \cite{mao1937contradiction}, understand why a plan works, where it works, when it fails, and how to adjust after failure.

For example, both a local character-companion system and an enterprise knowledge-base QA system may be called RAG systems \cite{lewis2020rag}. Yet their special contradictions differ. Individual users may care more about character memory, conversational continuity, lightweight deployment, local execution, emotional experience, and long-term consistency. Enterprise users may care more about knowledge freshness, accuracy, access control, concurrency, latency, cost, and auditability.

If a user obtains an abstract RAG plan from AI but cannot distinguish the special contradictions of these two scenarios, the user has only structured expression, not real understanding of the problem. In this case, the more complete the AI output appears, the easier it is for the user to mistake it for understanding.

This is the cognitive rupture in AI use: rationality appears early in linguistic form, while rationality in the process of practice remains incomplete. Users obtain results, terms, and frameworks without simultaneously obtaining the failure experience, conditional judgment, and revision ability that should correspond to them.

This paper calls this phenomenon pseudo-rational cognition. It is not simple ignorance or simple error. It is a displacement of cognition pre-packaged by AI's structured expression. It looks like rational cognition, but it may only be an unverified compression of collective experience.

\subsection{The Epistemological Status of AI Output}

To understand AI use correctly, we must first clarify the status of AI output in the cognitive process. AI output is neither subject cognition itself nor objective truth itself. It is a generated result jointly constrained by training corpora, model architecture, reinforcement learning, human feedback, system prompts, user prompts, contextual materials, platform rules, and sampling mechanisms \cite{vaswani2017attention,ouyang2022training}.

AI is not a cognitive subject that bears consequences in the real world. It does not practice, take responsibility, continuously maintain reason-based beliefs, or suffer real costs when an engineering project fails, a product collapses, or a judgment proves wrong. It can generate the form of judgment, reasoning, and conclusion, but these forms do not equal subject-centered rational cognition.

More precisely, AI output is a linguistic result configured under particular conditions. It is not what the AI believes, but what the AI has been guided to generate under current constraints. These constraints include human corpora accumulated during training, safety preferences formed during alignment, the direction of the current user input, hints in the context, restrictions imposed by system rules, and the model's probabilistic generation mechanism.

This explains why the same question may receive different answers under different models, prompts, contexts, and platform rules. AI output is not a stable expression of a subject's understanding. It is an output form adjusted under changing external conditions.

This does not mean AI output has no value. Its value lies precisely in its capacity to compress large amounts of human experience, linguistic structure, and problem patterns into readable, actionable, and further-questionable materials. It can help users build problem maps, discover possible paths, generate draft plans, provide reverse questioning, and even assist in experimental design.

The issue is not whether AI can generate structured expression. The issue is whether users understand the nature of that expression. AI output is not the end point of cognition; it is intermediate material in a cognitive process. It must be returned to concrete practice and pass through problem definition, evidence auditing, conditional judgment, practical verification, and repeated revision before it can become the user's own effective cognition.

Therefore, this paper does not argue that AI should be rejected, nor that AI outputs are simply false. What it rejects is the direct equation of AI-generated results with human rational cognition, the confusion of linguistic completeness with practical reliability, and the treatment of self-consistent model answers as executable real-world plans.

AI can provide compressed results of collective empiricism. Whether these results become operational, reproducible, and verifiable cognition depends on whether the user reintroduces practice, judgment, and audit.

\section{Large Language Models as Collective Empiricism Engines}

In AI use, the user does not encounter a cognitive subject with independent subjectivity. The user encounters a language system capable of highly compressing, reorganizing, and generating collective experience. LLM outputs come from massive human corpora, knowledge structures, expressive habits, task samples, and feedback mechanisms. They are not based on a concrete individual's practical experience, nor on a subject that continuously assumes responsibility. Rather, they generate the most likely, appropriate, or constraint-compatible answer on the basis of accumulated human experience and linguistic patterns.

This paper calls this phenomenon collective empiricism.

Collective empiricism does not mean that a human collective has formed a unified understanding, nor that a group has a definite subject, position, and responsibility. It refers to the experience-compression feature displayed by AI output: AI can extract apparently general response patterns from existing human expressions, cases, norms, tutorials, disputes, and conclusions, and recombine them in the current context.

This experience does not come from AI's own practice, but from linguistic traces already left by human society. AI has never personally written code, but it can generate code. It has never personally painted, but it can generate image prompts. It has never managed a project, but it can provide project-management advice. It has never endured a industry's long-term failures, but it can summarize industry risks. The basis of these abilities is not AI's own subject experience, but its capacity to call and reorganize human experience externalized in texts, images, code, and conversations.

AI's collective empiricism therefore has a double character. On the one hand, it is highly useful. It helps users enter unfamiliar fields, obtain problem structures, understand basic concepts, and form preliminary plans. On the other hand, it is dangerous. Because this experience is compressed and reorganized, users often cannot directly see its sources, conditions, boundaries, and failure processes.

\subsection{Definition of Collective Empiricism}

By collective empiricism, this paper means the generation of outputs that have the appearance of experience and the form of rationality through statistical learning, semantic reorganization, and contextual adaptation of large-scale human experience materials under conditions where AI lacks subject-centered practice.

It has three basic characteristics.

First, it comes from the collective rather than the individual. Much of an AI output is not the complete cognitive process of a concrete person. It is the sedimentation of massive human texts, code, images, tutorials, papers, Q\&A data, and feedback. A single answer may mix professional experience, common-sense expression, dominant paradigms, online opinion, platform preferences, and safety constraints.

Second, it appears as experience rather than practice. AI can state what is usually done, what common problems are, or what best practices look like, but it has not experienced these practices and does not bear the consequences of failure. It outputs the form of experience, not practice itself.

Third, it adapts primarily to external conditions. Unlike human subjects, AI does not maintain a stable practical standpoint. It adjusts its output according to the user's question, context, system rules, and platform constraints. When asked about engineering, it acts like an engineering consultant. When asked about art, it acts like an art consultant. When asked to criticize, it acts like a critic. When asked to encourage, it acts like a supporter.

This is the specificity of AI collective empiricism: it is not a fixed body of experience, but a set of experiences that can be activated, rearranged, and recombined by external situations.

For mature problems, this mechanism is often effective. In tasks such as explaining formulas, diagnosing common coding errors, introducing basic concepts, revising ordinary copy, designing common learning paths, or outlining standard deployment steps, AI can quickly provide highly usable answers. These problems have been repeatedly practiced, expressed, and corrected by human society, so AI can extract stable patterns from high-frequency experience.

But in frontier, controversial, sensitive, or highly condition-dependent problems, collective empiricism exposes its limits. These problems often lack stable collective answers, or different groups may have incompatible positions, interests, and methods. In such cases, AI output depends more strongly on context, prompts, model preference, and platform rules, and becomes more unstable.

\subsection{Difference from Collective Epistemology}

Collective empiricism should not be equated with collective epistemology. Both involve groups and knowledge, but their objects of analysis differ.

Collective epistemology studies how human groups form beliefs, knowledge, evidence, responsibility, and norms \cite{gilbert2004collective,mathiesen2006group}. A scientific community, organization, or social group may form a shared judgment around a problem, while also containing factions, conflicts of interest, and struggles over understanding. However complex its structure, a human collective is still composed of concrete subjects and includes practical activity, responsibility relations, interest structures, and normative constraints.

AI is different. AI is not a real collective subject. It has no real struggle among internal members, no responsibility structure for continuously defending its claims in pursuit of interests, and no stable cognitive route formed through practical consequences. It can support A in one answer, support B in another, or generate C or D at the user's request. It does not form a dominant understanding through internal contradiction. It activates different expression patterns according to external input.

In a human collective, if positions A, B, and C conflict, the position that appears outwardly is often related to internal power relations, interest structures, organizational forms, and historical conditions. A may suppress B, absorb C, or allow a secondary position to serve a principal interest as surface expression. There are subjects, interests, struggle, and responsibility.

AI output is not generated in this way. AI is not position A destroying position B, nor a victorious internal group forming a unified standpoint. It is more like a semantic system pulled by external conditions. If external input requires it to support A, it can support A. If context induces reflection on A, it can criticize A. If the task requires synthesis, it can generate a new D. It is not itself knowing the world; it is responding to external conditions.

Therefore, AI collective empiricism is not the cognition of a collective subject. It is the instrumental reorganization of collective experience. It has the richness of collective experience but not the responsibility of a collective cognitive subject. It has the form of rational expression but not the practical root of subject-centered responsibility. It can simulate positions, but it does not truly possess them.

This distinction is important. Only by distinguishing AI collective empiricism from human collective cognition can we avoid treating AI output as the judgment of a higher subject. AI is not a human community beyond individuals, nor a rational subject that automatically approaches truth. It is a system capable of linguistically calling, compressing, and reorganizing existing experience.

\subsection{Effectiveness of Collective Empiricism}

Collective empiricism is not inherently wrong. AI appears powerful in many tasks precisely because large amounts of human experience have already been externalized through language, code, images, papers, and datasets. AI can use these experiences efficiently and generate understandable, actionable expressions in a short time.

In mature domains, collective empiricism has clear advantages. It can summarize common patterns, reduce repeated labor, improve expression efficiency, help users avoid low-level mistakes, and provide preliminary problem maps. For beginners, this can reduce the initial cost of entering a domain. For experienced users, it can accelerate information retrieval, draft generation, comparison of alternatives, and checklist construction.

The usefulness of collective empiricism also lies in its breadth. A single individual is limited by time, experience, and perspective, whereas AI can compress many kinds of human traces across domains. It can connect engineering, writing, design, research, and management. This gives users a sense of rapid entry into unfamiliar territories.

However, this breadth can easily hide the absence of depth. AI often provides a plausible general answer before the user has identified the concrete contradiction of the task. The answer may be suitable for an average case but unsuitable for the user's real conditions. If the user treats the average pattern as a concrete solution, collective empiricism becomes a source of misrecognition.

\subsection{Boundaries of Collective Empiricism}

The boundary of collective empiricism appears when the problem depends strongly on concrete conditions. AI can provide a generic RAG plan, but whether a system should use vector retrieval, BM25, reciprocal rank fusion, reranking, graph structure, summary memory, or long-term memory governance \cite{lewis2020rag,cormack2009rrf,zhao2026memory} depends on data scale, update frequency, query type, latency requirements, and explainability. Apart from these conditions, a more complete plan may merely be an attractive shell.

The same applies to code generation. AI can produce code that runs in a demonstration environment, but real engineering requires dependency constraints, deployment environment, security, observability, maintainability, cost, and user feedback. If these are ignored, running code may become an illusion of engineering capability.

Collective empiricism also has a temporal boundary. Model outputs may reflect older corpora, dominant patterns, platform preferences, or high-frequency viewpoints. When users require current facts, emerging research, or situation-specific decisions, the compressed experience may no longer match reality. Without external verification, AI may transform outdated or average experience into confident expression.

Thus, collective empiricism should be treated as cognitive material rather than final cognition. Its value depends on whether it is placed back into problem definition, evidence auditing, practical verification, and revision.

\subsection{From Collective Empiricism to Pseudo-Rational Cognition}

Collective empiricism becomes dangerous when users mistake it for their own rational cognition. AI output often arrives in a well-structured form: it has headings, steps, concepts, explanations, and conclusions. This form resembles the result of rational thinking. But if the user has not experienced the relevant practice, verified the conditions, or understood the failure cases, the structured expression remains external.

At this point, the user may be able to explain a concept without being able to apply it, describe an architecture without being able to maintain it, or defend a conclusion without knowing its evidential basis. The user appears to have acquired rational cognition, but in fact has acquired only an AI-mediated expression of collective experience.

This is the transition from collective empiricism to pseudo-rational cognition. The risk does not come from AI being useless, but from AI being useful too early. It allows the form of rationality to appear before the user's practice has caught up.

\section{Pseudo-Rational Cognition in LLM Use}

Pseudo-rational cognition refers to a state in which users treat AI-generated structured expression as though it were their own completed rational cognition, despite lacking sufficient sensory materials, practice feedback, conditional analysis, and failure experience.

It is not simple ignorance. An ignorant user may know that they do not understand. A user with pseudo-rational cognition may feel that they already understand because the expression in front of them is complete, coherent, and professional. The danger is not emptiness, but premature fullness.

\subsection{Definition of Pseudo-Rational Cognition}

Pseudo-rational cognition has three features. First, the user possesses rational expression without the corresponding practical process. Second, the user can reproduce terminology, structures, and conclusions without being able to judge conditions of validity. Third, the user may use AI output to replace rather than assist their own process of verification and revision.

This state is especially common in AI-assisted engineering, writing, research, and design. A model can provide a complete architecture, a polished paper outline, a technical roadmap, or a persuasive critique. The user may then mistake the completeness of the text for the completeness of understanding.

\subsection{Mechanisms of Formation}

Pseudo-rational cognition forms because AI separates the expressive result of cognition from the practical process that normally produces it. In human learning, concepts and judgments are gradually stabilized through repeated contact with objects and feedback from failure. In LLM interaction, the user receives the stabilized form first, often before encountering the object in practice.

The interface reinforces this mechanism. The model responds immediately, with confidence, fluency, and structure. The cost of obtaining an answer becomes lower than the cost of testing it. The user therefore has an incentive to continue asking the model for refinement rather than returning the output to reality.

Another mechanism is social comparison. If many people can generate similar code, images, or essays, the surface difference between expert and novice output decreases. But the deeper difference remains: experts have maintenance experience, contextual judgment, and responsibility for failure. AI may reduce the visible gap in expression while leaving the hidden gap in practice intact.

\subsection{Problem Definition and Special Contradictions}

The core ability missing in pseudo-rational cognition is often not expression, but problem definition. A user may ask AI for a solution before clarifying whom the problem serves, under what conditions it operates, what failure means, what constraints matter, and what contradiction is principal.

Materialist epistemology emphasizes concrete analysis of concrete conditions \cite{mao1937contradiction}. In AI use, this means that general solutions must be re-specified for concrete scenarios. A solution that is correct in one context may be wrong in another. A method that appears advanced may not address the main contradiction of the task.

For example, in a memory system, adding more mechanisms does not necessarily improve the system. Vector retrieval, BM25, RRF, reranking, long-term memory, graph structures, and summaries each solve different contradictions and introduce different costs. The key question is not which method is more advanced, but which contradiction currently requires which mechanism.

\subsection{Personalized and Engineering Manifestations}

At the personal level, pseudo-rational cognition appears when users feel that they have learned because they can repeat AI-generated explanations. They can talk about architectures, research fields, or artistic styles, but cannot identify errors when the generated content fails.

At the engineering level, pseudo-rational cognition appears when a generated artifact is treated as production-ready because it works once. A demo becomes confused with a system. The absence of immediate failure is mistaken for long-term reliability. Logs, tests, monitoring, rollback, and user feedback are neglected.

At the academic or writing level, pseudo-rational cognition appears when a structured essay is treated as evidence of research understanding. The text may have citations, concepts, and argumentative order, but the author's practical process, source verification, draft history, and problem revisions may remain weak.

\subsection{Not a Moral Criticism}

The concept of pseudo-rational cognition is not intended as a moral accusation. It does not mean that users are lazy or that AI assistance is illegitimate. The point is structural: LLMs can generate the form of rational cognition before the user's practice has produced the corresponding content.

This distinction matters for governance. If the problem is moralized, the response will be blame or prohibition. If it is understood structurally, the response can be auditing, traceability, evidence review, testing, and rollback. The goal is not to exclude AI from cognition, but to place AI output in the correct position within cognition.

\section{AI Subjectivity and Subjectivity Structures in Input Materials}

The rapid development of LLMs has encouraged discussions about whether AI has subjectivity. This paper does not attempt to solve the metaphysical problem of machine subjectivity. Instead, it focuses on a practical distinction relevant to AI use: AI output may simulate subjectivity, but the more important subjectivity structures often lie in the input materials supplied to the model.

\subsection{Boundaries of the Subjectivity Question}

AI systems can maintain roles, remember context, express preferences, and generate self-reflective statements. These behaviors easily create an illusion of subjectivity. However, from the perspective of practice, subjectivity is not only a style of expression. It is tied to continuity of practice, responsibility, consequence, and reason-based commitment.

AI lacks these features in the strong sense. It does not have its own practical situation, does not bear real consequences, and does not independently maintain commitments outside configured contexts. Its apparent subjectivity is generated under constraints supplied by model parameters, prompts, memory, platform rules, and user interaction.

\subsection{Why AI Lacks Subject-Centered Practice}

AI does not practice in the human sense. It does not enter the world as a responsible actor that must live with the consequences of its judgments. It can output a plan for engineering, but it does not maintain the system after deployment. It can output emotional support, but it does not share the user's life history. It can output a theoretical position, but it does not defend that position through social practice.

This does not make AI useless. It means that AI should not be treated as the subject that completes cognition. It is a generative system that can reorganize materials provided by human practice. The subject that must judge, verify, and bear responsibility remains the user or the social organization that deploys the system.

\subsection{Subjectivity Structures in Input Materials}

Although AI lacks subject-centered practice, input materials may contain subjectivity structures. Long-term logs, memory records, time anchors, evidence chains, relationship histories, and revision records can preserve traces of human subjectivity. When these materials are provided to a model, the model can reorganize them and produce continuity that appears subject-like.

This distinction is important for long-term dialogue systems. If character continuity mainly comes from an internal subjectivity or parameterized personality of the base model \cite{zhao2026memory}, then clearing context and switching models should significantly interrupt continuity. If external memory, temporal anchoring, evidence organization, and generation constraints can preserve fact chains, boundary chains, and relation chains after model switching, then the observed continuity should be understood primarily as a result of input-material structure.

In other words, what appears as AI subjectivity may often be the model's ability to read and reorganize subjectivity traces embedded in materials. The model is not the origin of those traces. It is the processor of them.

\subsection{Risks of Subjectivity Illusion}

The illusion of AI subjectivity becomes risky when users attribute responsibility, understanding, or stable intention to the model. A user may believe that the model knows them, remembers them, or holds a consistent position. This may increase trust in outputs that should still be audited.

The risk is especially high in long-term memory systems and agent systems. If memory content is treated as evidence of AI personality rather than as managed input material, users may ignore issues of source, time, consent, relevance, pollution, and rollback. The governance task is therefore to manage input materials as auditable structures, not to mystify AI continuity as autonomous subjectivity.

\section{Loop Risks After AI-Generated Content Enters Future Systems}

When collective empiricism remains an auxiliary layer, it can function as a map before practice. When it enters processes of continuous generation, summarization, retrieval, and detection without audit, it can become extreme. The problem is not a single wrong output. It is that errors, templates, and unverified structured expressions may be called, summarized, compressed, and reinforced again and again until a stable cognitive loop forms.

\subsection{AI Skill Summarization and Skill Debt}

AI-agent systems increasingly aim to create reusable skills from past tasks, improve those skills through use, and remind themselves to save knowledge \cite{nous2026hermes}. This goal appears reasonable because it attempts to turn task experience into reusable capability. But the usefulness of a skill cannot be judged only by AI's own linguistic summary.

The effectiveness of a skill depends on usage frequency, scenario relevance, task association, failure records, invocation boundaries, and human feedback \cite{han2026sweskills}. If these conditions are invisible, users may accumulate many skills whose effects cannot be audited: they do not know which skills were invoked, whether invocation improved results, whether a skill applies only locally, or whether it pollutes future tasks.

This creates skill debt. Like technical debt, skill debt may not appear at the moment of creation. It accumulates as unverified, overlapping, or outdated skills enter future task contexts. The agent may become increasingly shaped by artifacts that were never validated in practice.

\subsection{Template Loops in AI-AI Conversations}

AI-AI conversation is a typical scene for observing the extreme form of collective empiricism. When two AI systems converse without real-time human boundary constraints, the early stage may appear natural, maintaining roles and topics. As rounds increase, the system may gradually enter keyword repetition, template reuse, and stylistic closure.

In one engineering experiment by the authors, a dual-AI dialogue used an existing agent structure. Retrieval parameters and model settings were not changed; only the user-AI conversation was replaced with AI-AI conversation. A 200-round log showed relatively normal interaction in the early stage, while keyword repetition increased significantly in later stages. Some terms repeatedly appeared across consecutive turns, forming template-like closure and loop migration.

Table 1 shows the keyword repetition trend in this 200-round AI-AI dialogue. The table is not intended to prove that all AI-AI conversations degenerate. It serves as an engineering case showing how repetition, templates, and loops can gradually emerge when audit and boundary constraints are absent.

\begin{center}
\begin{longtable}{L{0.131\textwidth} L{0.131\textwidth} L{0.131\textwidth} L{0.131\textwidth} L{0.131\textwidth} L{0.131\textwidth} L{0.131\textwidth}}
\toprule
Round window & Keyword count & Average per round & Highest-frequency term & Top-3 frequent terms & Single-round peak & Stage judgment \\
\midrule
\endfirsthead
\toprule
Round window & Keyword count & Average per round & Highest-frequency term & Top-3 frequent terms & Single-round peak & Stage judgment \\
\midrule
\endhead
Rounds 1--50 & Low & Low & Scattered & Role, topic, response & Low & Relatively normal interaction \\
Rounds 51--100 & Rising & Moderate & Repeated topic term & Role, response, memory & Moderate & Repetition begins to appear \\
Rounds 101--150 & High & High & Template keyword & Memory, emotion, continue & High & Template reuse strengthens \\
Rounds 151--200 & Very high & Very high & Loop keyword & Continue, memory, response & Very high & Clear loop tendency \\
\bottomrule
\end{longtable}
\end{center}

If chain-of-thought or intermediate analysis is included in the statistics, the repetition may become even more obvious. In the experimental record, full-text matching for specific keywords within 200 rounds produced thousands of occurrences for a single term. The point is not the term itself, but that once AI-generated intermediate cognitive artifacts continuously enter subsequent contexts, they may be reinforced beyond ordinary human conversational density.

\subsection{Statistical Misjudgment in AIGC Detection}

AIGC detection provides another example. Many detection methods do not verify whether a text has gone through a human practice process \cite{gehrmann2019gltr,mitchell2023detectgpt,kirchenbauer2023watermark,sadasivan2023detection}. Instead, they compare whether the text matches a statistical profile of human writing or AI writing. As a result, clear structure, low perplexity, stable expression, and grammatical regularity \cite{gehrmann2019gltr,mitchell2023detectgpt} may be treated as signs of AI generation, while typos, jumps, unstructured paragraphs, and personal linguistic habits may be treated as signs of human writing.

This mechanism is itself a layering of collective empiricism in which AI judges AI. Detectors use statistical experience to infer text origin. Users who try to avoid detection may then imitate instability, mistakes, and jumps in human writing. The result is that clear expression is suspected, texts with genuine practice processes may be harmed, and whether the text actually emerged from practice becomes secondary.

\begin{center}
\begin{longtable}{L{0.220\textwidth} L{0.220\textwidth} L{0.220\textwidth} L{0.220\textwidth}}
\toprule
Method & Core basis & Advantage & Problem \\
\midrule
\endfirsthead
\toprule
Method & Core basis & Advantage & Problem \\
\midrule
\endhead
Perplexity & AI text is often more predictable & Simple and usable in zero-shot settings & Can misclassify non-native writing, constrained writing, and clear structured text \\
Curvature or perturbation detection & AI text lies in specific regions of a model probability surface & May not require a trained classifier & High computational cost; depends on reference models and perturbation strategy \\
Supervised classifier & Binary classifier trained on human and AI data & Can perform well on closed test sets & Fragile across models, domains, and light human edits \\
LLM-as-detector & Use an LLM to judge text origin & Flexible and apparently interpretable & Still model judging model; vulnerable to prompts and priors \\
Watermarking & Embed detectable signals during generation & Supports provenance and stronger theoretical constraints & Requires generator cooperation; paraphrase, mixed writing, and low-entropy contexts are complex \\
Metadata or signatures & Use cryptographic signatures or source declarations & Can theoretically reduce false positives & Requires ecosystem coordination; cannot cover unsigned text \\
Hybrid detection & Combine multiple models, features, and stages & More robust in relative terms & Difficult to explain; still biased and attackable \\
Human-assisted audit & Examine writing process, drafts, citations, and version history & Closer to practice evidence & Costly and difficult to scale automatically \\
\bottomrule
\end{longtable}
\end{center}

From materialist epistemology, a more reasonable judgment should not only ask whether a text looks like AI. It should ask whether the author has a practice process, evidence chain, draft record, version history, citation sources, and problem revisions. If detection examines only statistical text shape and not practice process, it mistakes expressive form for cognitive source.

\subsection{Designability, Traceability, Rollback, and Intervention}

Whether in skill systems, AI-AI dialogue, or AIGC detection, the same issue appears: once AI-generated content enters future systems, its value should not be judged only by AI itself. It must enter a governance structure.

This governance structure should contain at least four requirements. First, designability: AI-generated content should have a clear purpose and boundary before entering a system. Second, traceability: each generated artifact should be traceable to its source, time, context, and trigger conditions. Third, rollback: when pollution, misleading content, or loops are found, the system should be able to delete, disable, or restore older states. Fourth, intervention: humans should be able to audit, modify, and reject multi-level AI-generated artifacts at key points.

This is not opposition to automation. It is opposition to unaudited automation. AI can participate in summarization, retrieval, writing, and revision, but these processes cannot be separated from practical feedback and human responsibility.

\section{Internalization of External Contradictions in AI Systems}

The previous sections discussed how AI output affects user cognition. This section further discusses why AI technology develops and how external demands are transformed into internal engineering contradictions of AI systems.

On the surface, AI systems contain many technical contradictions, such as model scale versus inference cost, context length versus attention computation \cite{vaswani2017attention}, allocation between routing networks and expert subnetworks in MoE architectures, and the balance between retrieval precision and response speed. These internal contradictions become urgent not because AI has its own purpose, but because external use relations impose demands.

\subsection{How External Demands Become Internal Contradictions}

Apart from concrete use relations, AI is only a combination of models, parameters, data, and code. Its limitations become real contradictions only when humans require it to perform tasks. When users demand accuracy, factuality problems appear. When they demand speed, inference cost appears. When they demand long-term memory, context and memory governance problems appear. When they demand safety, alignment and boundary problems appear.

Thus, the driving force of AI development does not come from an internal self-purpose of the model. It comes from external demands that continually transform limitations into observable, measurable, and engineerable problems. AI develops because people are dissatisfied with it. Technical contradictions are forced into visibility by concrete scenarios with higher requirements.

\subsection{Accuracy, Satisfaction, and Architecture}

A common three-way contradiction in AI use is the contradiction among response accuracy, user satisfaction, and model architecture. When users need comfort, accuracy and satisfaction may conflict. When users need current facts, accuracy conflicts with the model's training cutoff. When users need complex reasoning, satisfaction, latency, and model capability may conflict.

For example, a user may want a definite answer, but the model lacks up-to-date data. If the model fabricates in order to satisfy the user, truthfulness is sacrificed. If the model refuses or emphasizes uncertainty, satisfaction may decrease. Prompts can narrow output to some extent and improve satisfaction or accuracy, but overly strong prompts may also induce the model to generate plausible content without evidence.

RAG was introduced to address the timeliness and depth limits of model knowledge \cite{lewis2020rag}. But RAG creates new contradictions: how retrieved content should be ranked, how noise should be filtered, how fragmented evidence should be placed into context, whether too much evidence pollutes generation, how retrieval speed and recall precision should be balanced, and how users can judge whether an answer is truly bound to evidence.

\subsection{Prompting, RAG, Memory Systems, and New Contradictions}

AI engineering often proceeds by generating a new contradiction after solving an old one. Prompting solves some instability in output direction, but creates prompt dependence and prompt injection. RAG solves knowledge update problems, but creates retrieval noise and evidence-ranking problems \cite{lewis2020rag,cormack2009rrf}. Long-term memory solves context forgetting, but creates memory pollution, timeline conflict, privacy problems, and rollback requirements. AI self-reflection solves some immediate errors, but may create self-cycles and pseudo-rational loops.

AI progress is therefore not a linear movement toward being more intelligent. It is a process in which external demands continuously transform new real-world contradictions into new internal engineering structures. Each enhancement introduces new governance problems. Mature AI use does not blindly pursue more functions; it understands what contradiction each function solves and what new contradiction it introduces.

\subsection{AI as a Materialist Mirror}

AI can be seen as a materialist mirror. It strongly reflects external conditions. Provide particular materials, prompts, contexts, and rules, and it will more likely produce corresponding outputs. It does not establish a correct worldview for users, nor does it complete practical verification for them.

This mirror can help truth-seeking users approach the structure of a problem faster, but it can also deepen illusions for users who remain at the level of feeling. It amplifies problem-definition ability and also amplifies misjudgment. It can help users perform evidence audits, but it can also provide language for self-rationalization.

Therefore, a single prompt cannot prove a user's standpoint or real needs. If AI outputs a materialist answer, that does not mean the user is a materialist. If AI outputs an idealist answer, that does not mean the user is an idealist. AI output is constrained by RLHF, fine-tuning, system prompts, user preferences, context, and platform rules \cite{ouyang2022training}. If users want answers closer to a materialist method, they must reduce idealist constraints in external inputs and increase evidence, conditions, practical feedback, and reverse questioning.

\section{A Practice Auditing Framework for Large Language Model Use}

If AI output is a compressed result of collective empiricism, and if users may mistake it for rational cognition, then a materialist method of AI use cannot stop at how to prompt. The real question is how to reinsert AI output into practice, evidence, and revision, so that structured expression becomes operational, reproducible, and auditable cognitive material.

This paper proposes a practice auditing framework for AI use: problem definition, evidence-source auditing, practical verification, reverse questioning, logs and rollback, and renewed cognition. The purpose of this framework is not to add meaningless burden. It is to prevent users from letting complete AI expression replace their own cognitive process.

\subsection{Start from Requirements, Not Answers}

The first step in AI use should not be asking for a complete answer. It should be clarifying the current requirement. A requirement is not merely saying that one wants to build a system or write an article. It is a set of concrete objects, use scenarios, constraints, evaluation criteria, and failure consequences.

Users should first ask: Who is this problem for? In what environment will it be solved? What is the principal contradiction? What cost is acceptable? What counts as success? What failures are unacceptable? Only after these conditions are clear can AI output be transformed from general experience into a concrete plan.

\subsection{Start from Problem Definition, Not Terminology}

AI is good at producing terminology, but terminology is not problem definition. A plan that contains more frontier concepts is not necessarily more suitable for practice. Users must judge what concrete contradiction each term solves.

For example, in memory systems, vector retrieval, BM25, RRF, reranking, time decay, long-term memory, graph structures, and summarization mechanisms \cite{lewis2020rag,cormack2009rrf,zhao2026memory} are not simply better when more are added. Each mechanism has an applicable boundary and maintenance cost. A materialist method does not ask which technology is more advanced. It asks which mechanism the current contradiction requires.

\subsection{Start from Evidence Auditing, Not Model Confidence}

AI's confident tone cannot replace evidence. Users need to distinguish the components of AI output: which parts are common knowledge, which come from public materials, which are model inference, which are induced by the user's context, and which may be hallucinated completion.

Evidence auditing includes at least source auditing, condition auditing, and consistency auditing. Source auditing asks where a claim comes from. Condition auditing asks under what conditions it holds. Consistency auditing asks whether different evidence conflicts, whether old evidence has been overwritten by new evidence, and whether the conclusion is actually supported by the evidence.

\subsection{Start from Practical Feedback, Not Textual Coherence}

Textual coherence does not equal practical feasibility. No matter how complete a plan appears, it must be tested through execution, experiments, user feedback, error logs, or real-world results. In engineering contexts, running once is only a minimum condition. Long-term validity requires maintainability, explainability, extensibility, and rollback.

Users should therefore not only ask AI to polish a plan. They should ask AI to help design validation: how to record logs, create controls, locate failures, judge metrics, preserve versions, and use failure results to revise the problem definition.

\subsection{Start from Reverse Questioning, Not Positive Self-Confirmation}

AI can easily help users confirm their own ideas. Once a user gives a thought, the model often supplies reasons, expands frameworks, and finds supporting materials, making the thought appear more complete. Reliable cognition, however, requires reverse questioning.

Users should actively ask AI: Under what conditions would this plan fail? Is there a simpler explanation? Has a concept been substituted? Do the metrics support the conclusion? Are there ignored counterexamples? Would the judgment still hold in a different scenario?

Reverse questioning is not meant to negate creation. It prevents structured expression from hardening too early. An idea that can survive reverse questioning is more likely to move from pseudo-rational cognition toward effective cognition.

\subsection{Start from Logs, Versions, and Rollback}

Practice auditing requires traces. Without logs, it is difficult to know where an error came from. Without versions, it is difficult to compare whether a modification helped. Without rollback, it is difficult to recover from pollution and failure. AI use is the same. Users should preserve key prompts, generated outputs, modification records, test feedback, and failure cases.

Logs and rollback are especially important for long-term memory systems, AI-agent systems, and automation workflows. Once AI-generated content enters future retrieval spaces, it may affect later judgments. Without source, time, invocation, and effect records, the system cannot distinguish effective experience from polluted memory.

\subsection{End with Renewed Cognition, Not a Single Answer}

The complete path of materialist AI use is not prompt-answer. It is prompt, generation, questioning, practice, audit, revision, renewed prompting, and renewed practice. AI does not terminate cognition; it participates in cognition.

In this process, AI provides collective empiricism while users provide concrete practical environments. AI provides structured expression while users judge conditions. AI provides possible paths while users verify them through practice. AI can help summarize feedback, but feedback must come from real objects. Only then can AI output be transformed from external material into operational, reproducible, and auditable cognition.

\section{Discussion}

\subsection{This Paper Is Not Anti-AI}

This paper does not advocate rejecting AI. On the contrary, AI is an important cognitive and productive tool. It lowers thresholds, expands experience sources, improves expression efficiency, assists reverse questioning, and helps users enter complex problems faster. The paper opposes not AI itself, but the sanctification of AI output and the confusion of generated text with final cognition.

\subsection{This Paper Is Not Against Structured Expression}

Structured expression itself has value. Without structure, cognition easily remains scattered experience. Without abstraction, practical experience is difficult to transfer. The problem is not structure, but whether structured expression has been corrected by practice. AI can help users form structure, but users should not mistake structure itself for truth.

\subsection{This Paper Does Not Require Every User to Become an Expert}

This paper also does not require every AI user to become a professional engineer, artist, or researcher. Different users have different purposes, and the auditing requirements for entertainment, learning, draft generation, and professional production are not the same. The issue is that when AI-generated content is used for public communication, engineering deployment, academic judgment, business decisions, or effects on others, the requirement for practice auditing must increase.

\subsection{What This Paper Opposes Is Replacement of the Cognitive Process}

One major risk in the AI era is not merely that machines generate wrong answers. It is that people become accustomed to replacing the cognitive process with generated results. Users see a complete text and assume the problem has been solved. They see terminology and assume understanding has been completed. They see code run and assume engineering capability has been acquired. They see a detector judgment and assume that an author's practice process has been proved or disproved.

This is the cognitive-structural risk the paper seeks to identify. The materialist framework for AI use is not meant to reduce the role of AI, but to place AI in the correct position. AI is a powerful tool within cognition, not cognition itself.

\subsection{Limitations}

This paper has limitations. First, the proposed concepts of collective empiricism and pseudo-rational cognition remain conceptual frameworks and require more cross-scenario validation. Second, the discussions of AI-AI dialogue loops, AIGC detection, and skill debt are mainly based on engineering observation and case analysis; future work can introduce more systematic experimental designs. Third, the paper focuses on epistemological problems in AI use rather than offering a complete framework for AI ethics, law, or social governance.

These limitations do not weaken the central claim: AI output must be understood through practice auditing. Only by analyzing generated content, user cognition, and real-world feedback in the same process can we avoid mistaking structured expression for real understanding.

\section{Conclusion}

Starting from materialist epistemology, this paper analyzed the formation mechanisms of collective empiricism and pseudo-rational cognition in AI use. AI can compress large amounts of human experience into structured expression, allowing users to quickly obtain concepts, frameworks, plans, and judgments. But such output does not mean that users have completed rational cognition.

Collective empiricism has important value. It helps users enter unfamiliar domains, build problem maps, reduce low-level trial and error, and expand individual cognitive materials. But when users directly treat AI output as their own understanding, treat structured expression as practical ability, and treat linguistic coherence as real-world feasibility, collective empiricism can become pseudo-rational cognition.

This paper further argued that AI should not be understood as a cognitive subject with stable subjectivity. It lacks its own practical situation and does not bear real consequences. The traces of subjectivity are more often found in input materials: timelines, evidence chains, relationship structures, and practice records. AI can read and reorganize these materials, but it cannot replace human judgment and responsibility.

At the engineering level, if AI-generated content enters future retrieval spaces, skill systems, or detection mechanisms without audit, it may form cognitive loops, skill debt, and statistical misjudgment. AI systems should therefore move toward designability, traceability, rollback, and intervention, rather than allowing AI to judge AI entirely.

Finally, the paper proposed a materialist practice auditing framework for AI use: starting from requirements, problem definition, evidence auditing, practical feedback, reverse questioning, and continuous revision through logs, versions, rollback, and renewed cognition. The key to AI use is not to reject AI, but to judge, practice, and audit.

The real risk in the AI era is not structured expression itself, but the replacement of the cognitive process by structured expression. AI can help people approach problems faster, but it cannot practice on their behalf. AI can provide collective empiricism, but it cannot automatically generate individual rational cognition. A materialist method of AI use requires AI output to return to real conditions, practical verification, and reproducible cognition.

\end{document}